\documentclass[aps,prb,reprint,superscriptaddress]{revtex4-1}

\usepackage{graphicx}
\usepackage{color}

\def\kp{$k\cdot p\ $}
\def\changed#1{{\color{black}#1}}
\begin{document}

\title{Sensitivity of the MnTe valence band to orientation of magnetic
moments} %: effective model and QSGW calculations}

\author{Paulo E. Faria Junior}
\affiliation{Institute for Theoretical Physics, University of
  Regensburg, D--93040 Regensburg, Germany} 

\author{Koen A. de Mare}
\affiliation{Eindhoven University of Technology, Eindhoven
NL--5612 AZ, Netherlands}
\affiliation{FZU -- Institute of Physics of the Czech Academy of 
Sciences, Cukrovarnick\'a 10, Praha 6, CZ--16253}

\author{Klaus Zollner}
\affiliation{Institute for Theoretical Physics, University of
  Regensburg, D--93040 Regensburg, Germany} 

\author{Kyo-hoon Ahn}
\affiliation{FZU -- Institute of Physics of the Czech Academy of 
Sciences, Cukrovarnick\'a 10, Praha 6, CZ--16253}

\author{Sigurdur~I.~Erlingsson}
\affiliation{Department of Engineering, School of Technology,
  Reykjavik University, Menntavegi 1, IS-102 Reykjavik, Iceland}

\author{Mark van Schilfgaarde}
\affiliation{National Renewable Energy Laboratory, Golden, CO 80401, USA}
\affiliation{King's College London, Physics Department, Strand, London WC2R 2LS, United Kingdom}

\author{Karel V\'yborn\'y}
\affiliation{FZU -- Institute of Physics of the Czech Academy of 
Sciences, Cukrovarnick\'a 10, Praha 6, CZ--16253}

\date{Feb14, 2022}   

\begin{abstract}
An effective model of the hexagonal (NiAs-structure) manganese telluride 
valence band in the vicinity of the A-point of the Brillouin zone is
derived. It is shown that while for the usual antiferromagnetic order
(magnetic moments in the basal plane) band splitting at A is small,
their out-of-plane rotation enhances the splitting dramatically (to
about 0.5~eV). We propose extensions of recent experiments (Moseley et al.,
Phys. Rev. Materials 6, 014404) where such inversion of
magnetocrystalline anisotropy has been observed in Li-doped MnTe, to
confirm this unusual sensitivity of a semiconductor band structure to
magnetic order.
%Such experimental data could also resolve the controversy about the competing valence band maxima.  location of the VB top: A or $\Gamma$.
\end{abstract}

\pacs{later}

\maketitle

\section{Introduction}

The electronic structure of crystalline semiconductors can be treated by
various methods which differ greatly in their computational
cost.\cite{someBook} Among {\em ab initio} methods, GW is one of the
most advanced approaches %(good gaps, PRL 99, 246403) which is, alas,
yet a numerically rather expensive one.\cite{Grumet:2018_a}
A widely-used alternative is density functional theory (DFT)
where the speed comes at the cost of worse performance (even if there
are various approaches to mitigate deficiencies such as too small gaps)
%lower reliability (bad gaps but good$a_{lat}$,Stefaan\cite{Lejaeghere:2016_a})
and yet faster options are available, of which tight-binding
approaches\cite{Andersen:1984_a} and \kp models\cite{Winkler2003}
will be of interest here. Such effective models need material
parameters (such as on-site energies or hopping amplitudes) as an
input which can sometimes be of advantage because they can be adjusted
to fit experiments. Also, they may offer insight into mechanisms
governing the band structure.
%such as level splittings sensitive to
%the orientation of magnetic moments in antiferromagnetic materials
%which will be of interest in the present work.

An archetypal example of an effective model is the Kohn-Luttinger
Hamiltonian\cite{Luttinger:1956_a} which has a wide range of
applications to non-magnetic materials, including silicon and III-V
semiconductors with $\Gamma_8$ manifold at the top of the valence band
(VB). Magnetism adds a new twist: for Mn-doped GaAs, the host is
described by this Hamiltonian and the effect of ferromagnetic
ordering is captured by a kinetic $pd$ exchange term
$\propto\hat{\vec{s}}\cdot\vec{S}$ where $\hat{\vec{s}}$ is the spin
operator (of the VB holes) and $\vec{S}$ is the classical spin
representing the Mn magnetic moments (usually treated on the
mean-field level). Such description of ferromagnetic
semiconductors\cite{Tanaka:2021_a,Jungwirth:2014_a} has been employed
extensively in the context of spintronics\cite{Marrows:2011_a} and
now that {\em antiferromagnetic} spintronics\cite{Baltz:2018_a} has
become an active field, we hereby wish to contribute to its progress by
presenting an effective model of hexagonal (NiAs-structure) MnTe
which is a well-established antiferromagnetic semiconductor, as
exemplified by its $T=0$ band structure in Fig.~1, with a
relatively high ($\approx 310$~K) N\'eel temperature. Typical samples,
both bulk and layers exhibit p-type conductivity and we will therefore
focus on its VB. %\changed{[\onlinecite{Bossini:2020_a}]}

%It is clear from the onset that $\propto\hat{\vec{s}}\cdot\vec{S}$ is
%not going to do the trick with MnTe. This additional term is odd in
%$\vec{S}$ and therefore, an antiferromagnetic system must be treated
%differently... 

\begin{figure}
  \includegraphics[scale=0.3]{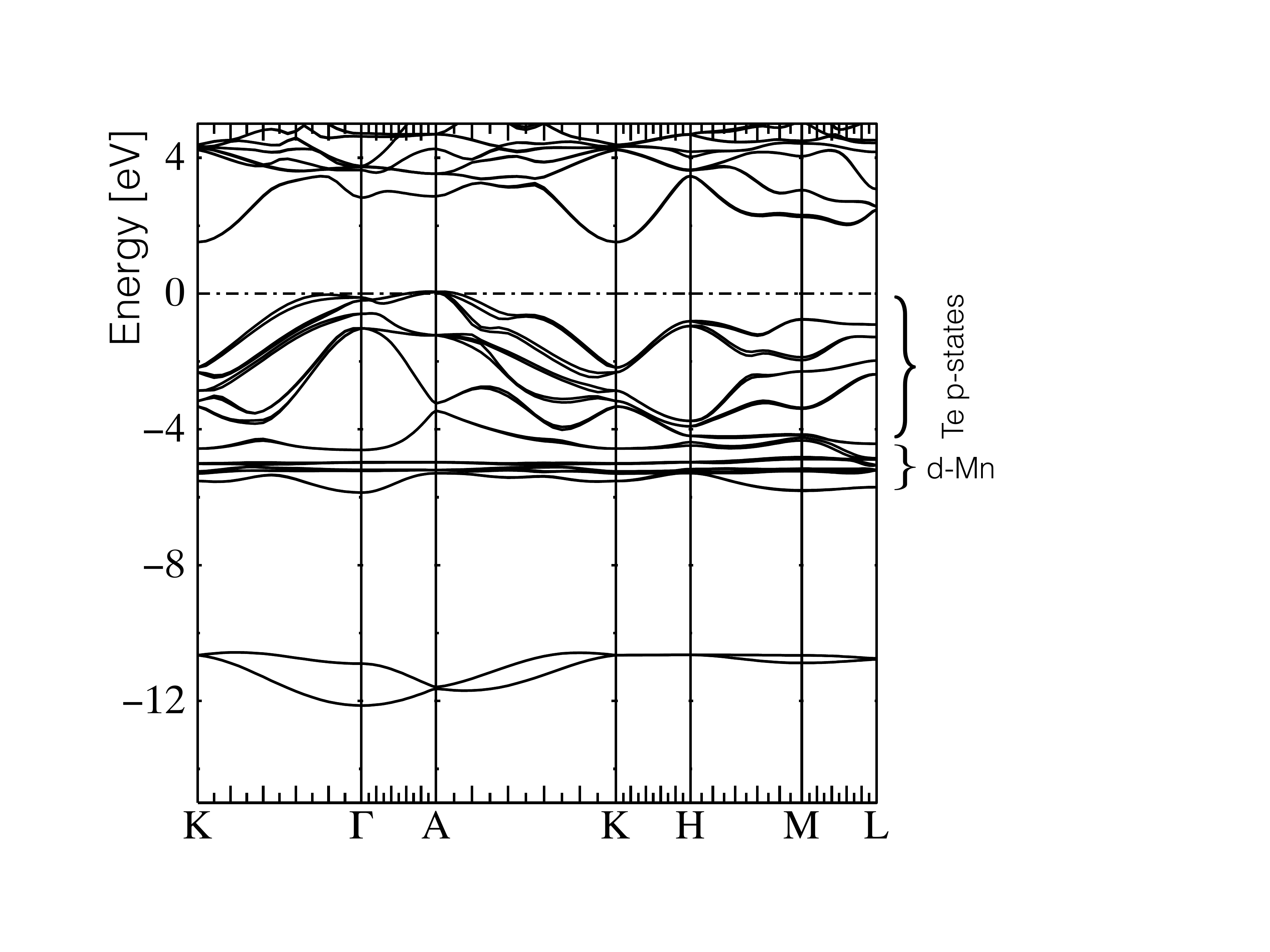}
  \vskip-1cm
  \caption{Band structure of MnTe calculated by QSGW 
    for $\vec{L}\perp z$ (in-plane orientation). Note the
    competing maxima of the valence band at $\Gamma$ and $A$ points.}
  \label{fig-01}
\end{figure}

The magnetic structure of MnTe was established\cite{Komatsubara:1963_a}
long ago (see Fig.~3) with a strong anisotropy favouring in-plane
orientation of the magnetic moments and a weak residual anisotropy
within the plane.\cite{Kriegner:2017_a}  Recently, Moseley et
al.~\cite{Moseley:2022_a} have found by neutron diffraction that, upon
doping by lithium, the magnetic moments rotate out of plane. They
also noticed that in the density of states (DOS), significant
changes occur and we use the effective model to explain how the VB
responds to this change of magnetic order (once spin-orbit interaction
is taken into account). Even if the Mn $d$-states lie\cite{Sato:1994_a}
deep below the Fermi level $E_F$ and seem too remote from the
VB top\cite{Bossini:2020_a}  which is built dominantly from $p$-Te
orbitals, we demonstrate that the combination of MnTe layered
structure and relativistic spin-orbit interaction (SOI) lead to
an unusual sensitivity of the electronic structure to the orientation
of magnetic moments. In the next Section we discuss the competing VB
maxima and we focus on the one near A-point of the Brillouin zone (BZ)
in Sec.~III. We conclude in Sec.~IV.

\section{Competing VB maxima}  % \onlinecite{Yin:2019_a}}

Once the SOI is taken into account, there arises a tight competition
between valence band maxima close to A and $\Gamma$ points of the BZ,
see Fig.~2b. A long-standing consensus\cite{Podgorny:1983_a,note3} that
the former prevails has recently been challenged by
Yin et al.~[\onlinecite{Yin:2019_a}] who claim that the VB top
occurs in the vicinity of $\Gamma$ point. To improve on the potentially
less accurate DFT approach,\cite{Yin:2019_a} we employ the
Quasiparticle Self-Consistent GW approximation\cite{Kotani:2007_a} (QSGW).
The GW approximation, which is an explicit theory of excited states is
widely used to predict quasiparticle levels with better reliability than
density functionals.  QSGW is an optimized form of the GW approximation,
where the starting Hamiltonian is generated within the GW approximation
itself, constructed so that it minimizes the difference between the
one-body and many-body Hamiltonians.  As a by-product the poles of the
one-body Green's function coincide with the poles of the interacting
one: thus energy band structures have physical interpretation as
quasiparticle levels, in marked contrast to DFT approaches \changed{(some
examples are given in Sec. 5.1 of the Supplementary Information)} where the
auxillary Hamiltonian has no formal physical meaning (in practice
Lagrange multipliers of this Hamiltonian are interpreted as
quasiparticle levels). In practice QSGW yields high fidelity
quasiparticle levels in most materials where dynamical spin fluctuations
are not strong.\cite{Schilfgaarde:2006_a} % and more refs. from 19omm/koresp/12

Bulk lattice constants of MnTe at room temperature are $a=0.414$~nm
and $c=0.671$~nm;\cite{Moseley:2022_a} % perhaps better cite [33] in 17nmt?
we show in Fig.~2d that for such $c/a=1.621$, the VB maximum close to
the A point safely prevails ($\Delta E$ is the difference between
energy of local VB maxima close to $\Gamma$ and that close to~$A$).
Most experiments nowadays are performed with
thin films of MnTe, however, and then lattice constants depend on the
choice of substrate. Temperature-dependent data in Fig.~3 of
Ref.~\onlinecite{Kriegner:2017_a} suggest that while samples grown on SrF$_2$
surface still fall into the same class, low temperatures may effectively
push the VB maxima close to the $\Gamma$ point up and in particular,
samples grown on the InP substrate may exhibit the inverted alignment of
the VB maxima.

Comparing the present QSGW results to DFT
calculations of Ref.~\onlinecite{Yin:2019_a}, several remarks are in
order. Lattice constants used in that reference (which correspond to
$c/a=1.57$) have been obtained by structure optimisation in DFT
rather than from experimental data. Next, the hybrid functional HSE06
may avoid the known problem of underestimated gaps in DFT but this in
itself does not guarantee a reliable description of finer details of
the band structure (such as VB maxima alignment). Predicted valence
and conduction bands are more uniformly reliable in GW than in
density-functional methods. Moreover QSGW surmounts the problematic
starting-point dependence that plagues the usual implementations of the GW
approximation % more refs. from 19omm/koresp/12 ?
and therefore QSGW is a better choice for our study than DFT.
Regarding the subsequent derivation of an effective model for the VB
around $\Gamma$ point,\cite{Yin:2019_a} we note as follows.
The $k_z=0$ approximation is used; while this would be appropriate for very
thin layers (say 5~nm), present experiments~\cite{Kriegner:2017_a} are more
likely behaving like 3D bulk. Also, the effective model~(1) in
Ref.~\onlinecite{Yin:2019_a} assumes a fixed direction of the magnetic
moments; to plot the experimentally relevant 'angular sweeps', the
current direction rather than N\'eel vector is rotated which is,
however, not the actual experimental protocol. For systems where only
the non-crystalline anisotropic magnetoresistance (AMR)
occurs,\cite{Rushforth:2007_a} the two protocols are equivalent but
measurements in the Corbino geometry\cite{Kriegner:2017_a} prove this
assumption false.  Being aware of these issues, we strive to derive
an effective model in the following which is free of these shortcomings captures the dependence on
magnetic moments direction.

\begin{figure}
  \begin{tabular}{cc}
    (a) & (b) \\[-2cm]
    \includegraphics[scale=0.22]{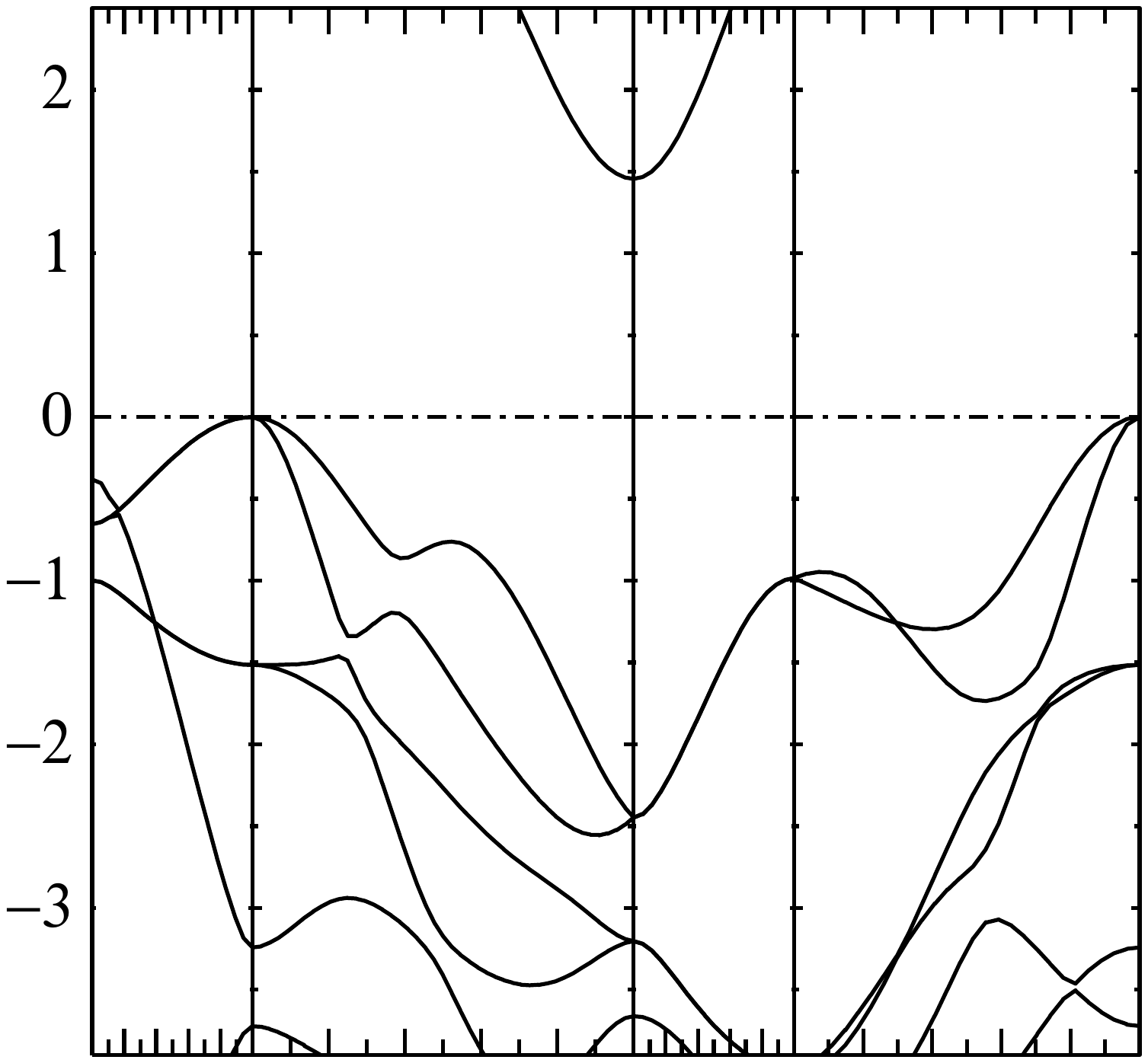}
    \put(-100,22){$\Gamma$}
    \put(-90,22){{\footnotesize A}}
    \put(-58,22){\footnotesize K}
    \put(-44,22){\footnotesize H}
    \put(-15,22){\footnotesize A}
    &
    \kern-20pt\includegraphics[scale=0.22]{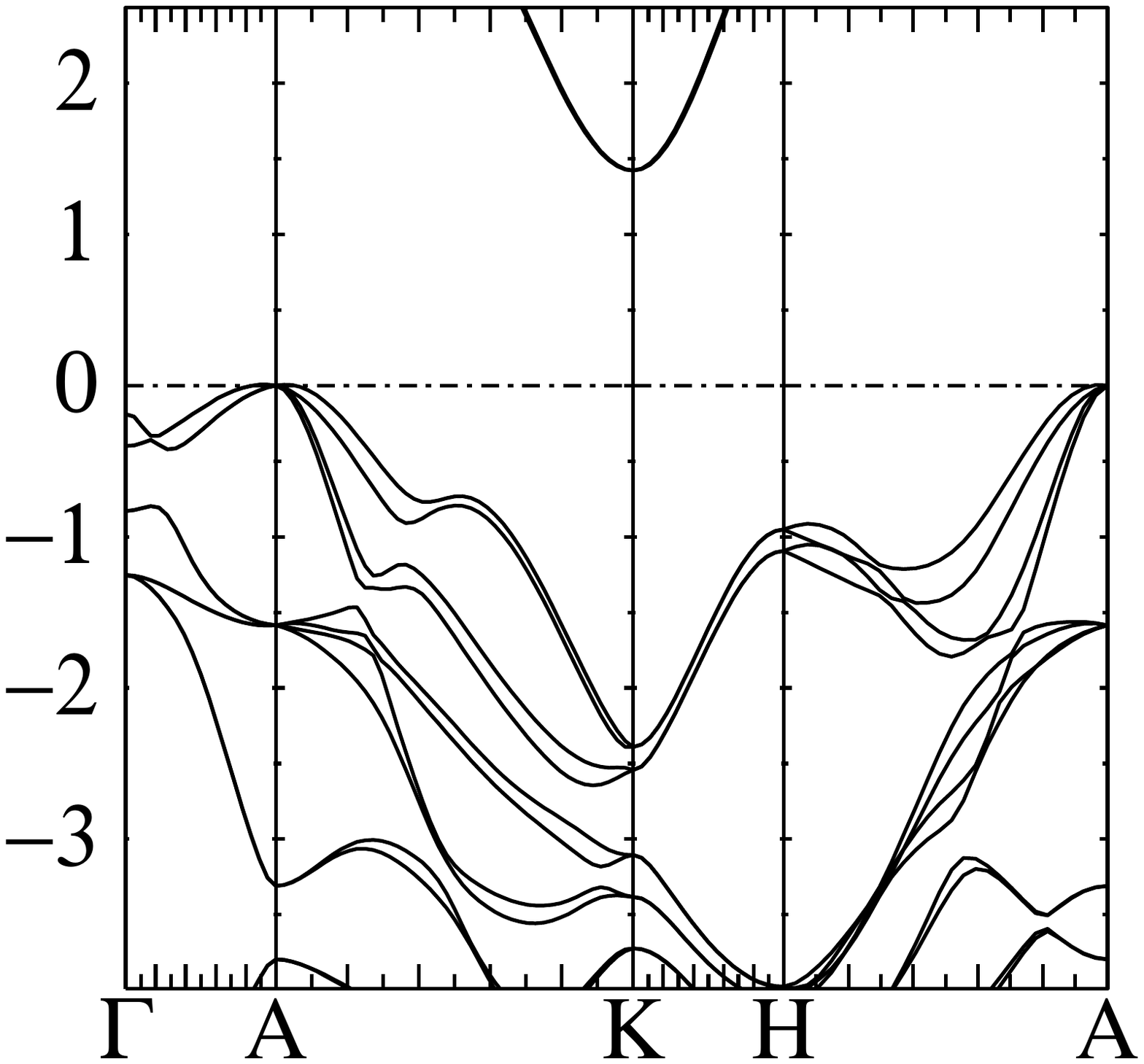} \\
    (c) %cryst. str
    &  (d) %Mark's plot
    \\
    \includegraphics[scale=0.1]{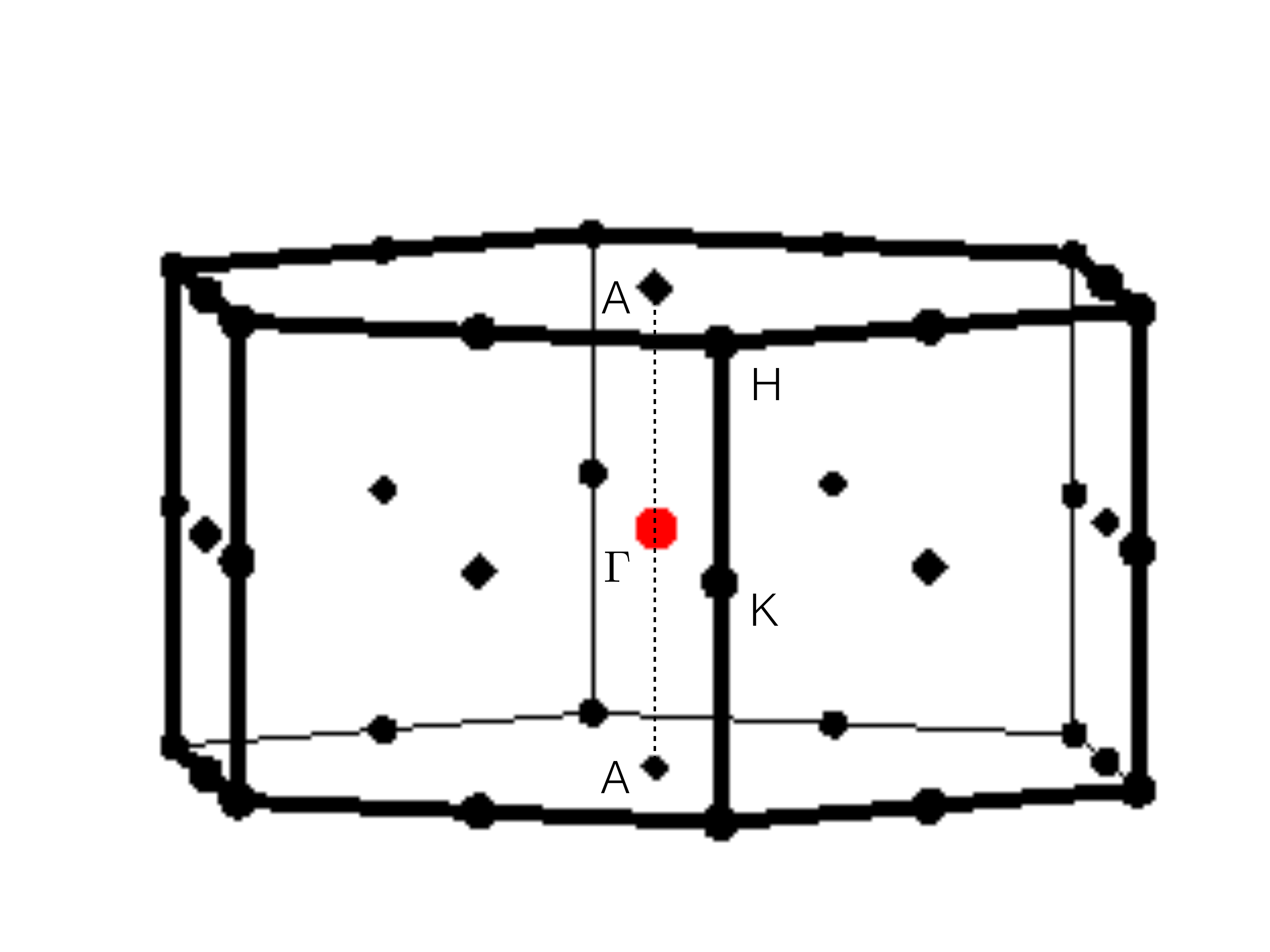} &
    \includegraphics[scale=0.22]{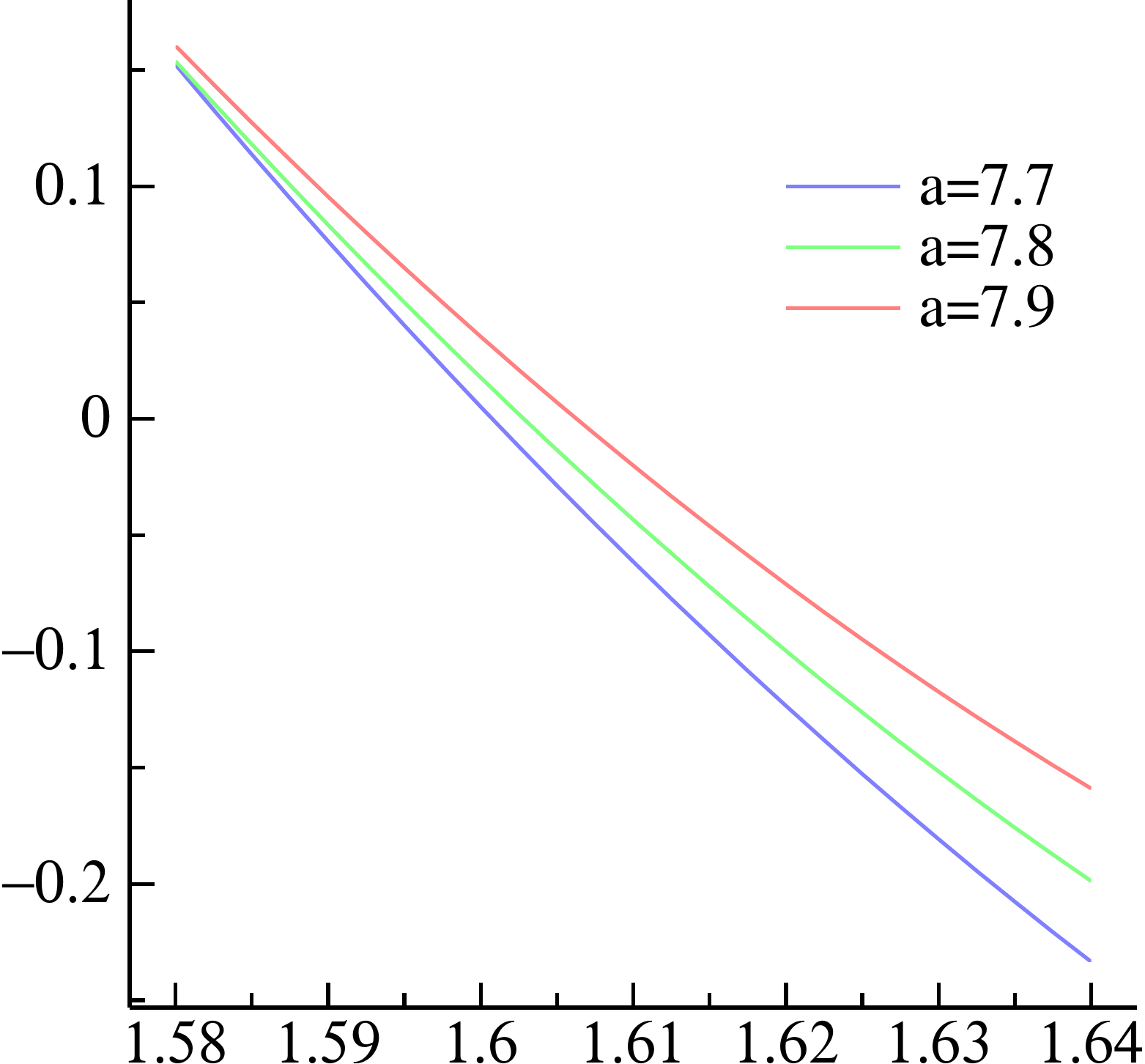}
  \end{tabular}
  \caption{Detailed view of the VB maximum and (c) the Brillouin 
    zone of $\alpha$-MnTe corresponding to lattice constants: 0.414~nm
    and $c/a=1.6208$. Panels (a,b) show the difference between SOI ignored
    and included and (d) gives for the latter case the energy difference
    between VB maximum in $\Gamma$ and A depending on the $c/a$ ratio
    (for several values of $a$ given in a.u.);  \changed{origin of this
    dependence is discussed in Sec.~5.2 of the Supplementary Material.}
    Negative $\Delta E$ means
    that the VB maxima around A prevail and all energies are given in eV.}
    %{\em Note:} 0.414~nm corresponds to 7.81~a.u.
  \label{fig-02}
\end{figure}

A proper symmetry analysis of the crystal structure of MnTe provides 
the non-symmorphic space group D$_{\text{6h}}^{\text{4}}$. Once AF ordering 
is included the Mn atoms must be treated as inequivalent since each Mn layer would have spins pointing in the opposite direction as shown in Fig.~\ref{fig-03} for in-plane spins. Hence, the symmetry group is reduced from D$_{\text{6h}}$ to D$_{\text{3d}}$ without SOI (see for instance Sandratskii et al.\cite{Sandratskii:1981_a}). Furthermore, the symmetry group would also depend  on the interplay of SOI and choice of the AF direction since spins pointing in different directions behave differently under symmetry operations. For example, in the out-of-plane AF configuration, the symmetry remains D$_{\text{3d}}$ while for in-plane AF, either along $\left[10\bar{1}0\right]$ or $\left[11\bar{2}0\right]$ directions, the symmetry group is reduced C$_{\text{2h}}$. Besides the conceptual analysis of the symmetries, independent calculations using the WIEN2k and Quantum Espresso ab initio packages also provide the same symmetry groups discussed above. Thus, for the particular choice of in-plane AF the D$_{\text{2h}}$ point group discussed by Yin et al. should be replaced by C$_{\text{2h}}$.

\section{Effective models}

%Apart from the recent study of Yin et al.\cite{Yin:2019_a},

Several attempts to describe the electronic structure of $\alpha$-MnTe
in a simplified way have been made so far. Here, the $k\cdot p$
approach\cite{Winkler2003,LewYanVoon2009} 
is a common choice for semiconductors\cite{FariaJunior2016PRB} especially
if only high-symmetry points in the BZ are of interest. Such a model for the
VB top in A point was derived more than 40 years ago\cite{Sandratskii:1981_a}
and later extended to a tight-binding scheme.\cite{Masek:1987_a} The latter
allows for the description of the energy bands over the whole BZ but neither
of these models allows to analyse the dependence of electronic structure
on the directions on Mn magnetic moments. In the perfectly ordered AFM
phase (as in Fig.~\ref{fig-02}d) and without SOI, the Bloch functions
at the top of the valence band in the A  point transform as the
two-dimensional irreducible representation E$_{\text{g}}$ or ($\Gamma_3^+$)
of the the D$_{3d}$. Including corrections up $k^2$ and no SOI
(essentially given by Eq.~2 in Sandratskii~et~al.\cite{Sandratskii:1981_a})
one would obtain the following Hamiltonian:
\begin{equation}
  H_{kp,2\times 2}=\left(\begin{array}{cc}
    ak_x^2+bk_y^2+ck_z^2 & (a-b)k_xk_y \\
    (a-b)k_xk_y & bk_x^2+ak_y^2+ck_z^2
    \end{array}
  \right) \, .
  \label{eq-01}
\end{equation}
%
%The  parameters\cite{note2} $a,b,c$ (in units of $\hbar^2/2m_0$) correspond to 
%(anisotropic) effective masses and can be fitted to the ab initio calculations 
%in Fig.~\ref{fig-02}(a).
The inverse effective masses (proportional to $a,b,c$) imply that Fermi
surfaces (FSs) are, at this level of approximation, two prolate ellipsoids
(both doubly degenerate) touching at the point where they are pierced
by the A$\Gamma$ line; %note that while MnTe is a hexagonal crystal, the
%model~(\ref{eq-01}) has a continuous rotation symmetry around the $z$-axis.
\changed{other properties of this model and its parameters, effective
  masses, extracted from fits to QSGW are given in Sec. 2 of the
  Supplementary Material.}
From the point of view of magnetism, this is a consequence of neglecting the
spin-orbit interaction. Once SOI is included, the band dispersion will
depend on the direction of magnetic moments. On the other hand, if higher
order terms in $\vec{k}$ were included, the symmetry would be lowered and
FSs would become warped and spin-split.\cite{Smejkal:2021_a}  Consequent
spin order in reciprocal space, being a hallmark\cite{Mazin:2021_a} of
so called altermagnetism, can lead to phenomena normally unexpected in
collinear compensated magnets such as the anomalous Hall effect.
%When $k_z=0$, one of the eigenstates of
%this matrix will be parallel to $\vec{k}$ while the other will be
%orthogonal to $\vec{k}$.

The derivation of~Eq.~(\ref{eq-01}) is based solely on symmetry arguments and
entails neither any explicit information about orbital composition of the
corresponding Bloch states nor any parametric dependence on magnetic
order. %The latter is, however, one of the necessary ingredients of the
%effective model, should it be capable of fully describing the AMR
%(which may be a manifestation of band structure sensitivity to the
%direction of magnetic moments).
In the following, we therefore first
describe a toy model capturing the essence of interplay between
magnetism and orbital degrees of freedom and next, we make use of these
insights to derive a realistic model of MnTe.

\subsection{Toy model}

\begin{figure}
  \includegraphics[scale=0.33]{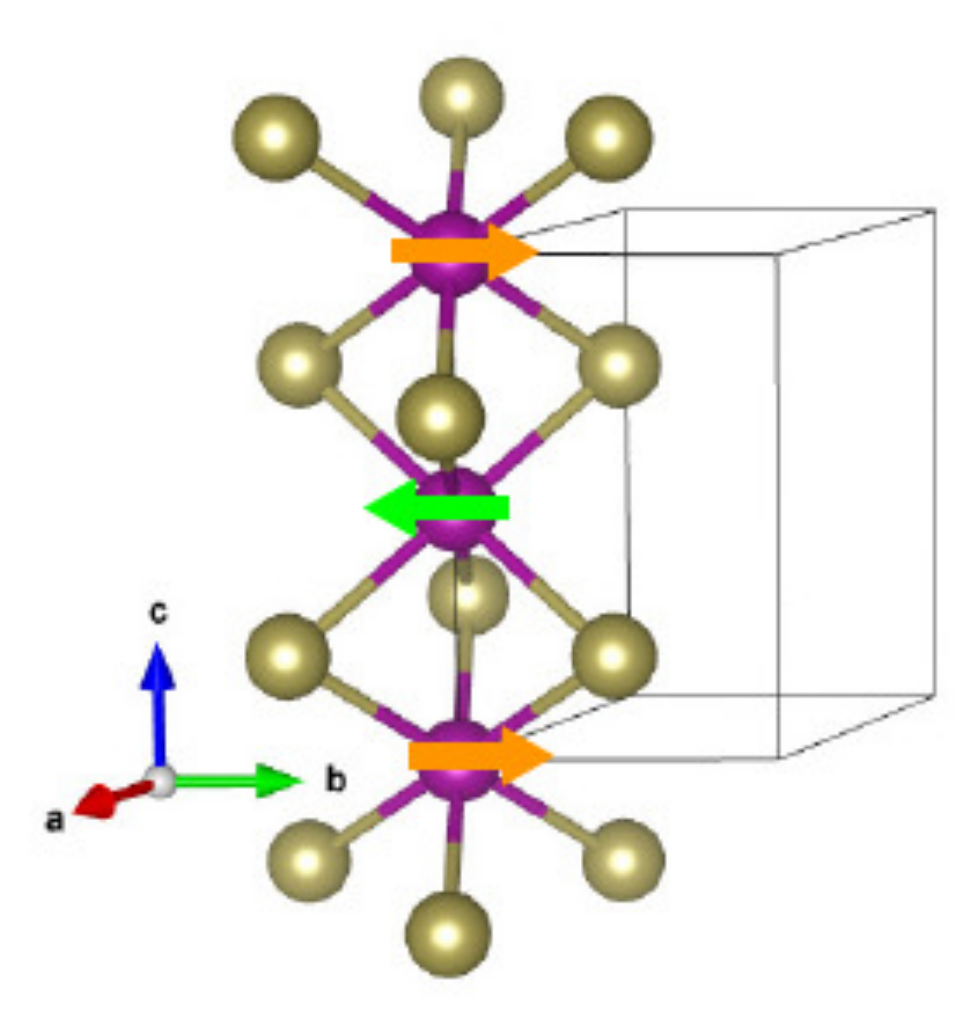}
  % used convertio.co on .ppm (obtained from png)
  \includegraphics[scale=0.17]{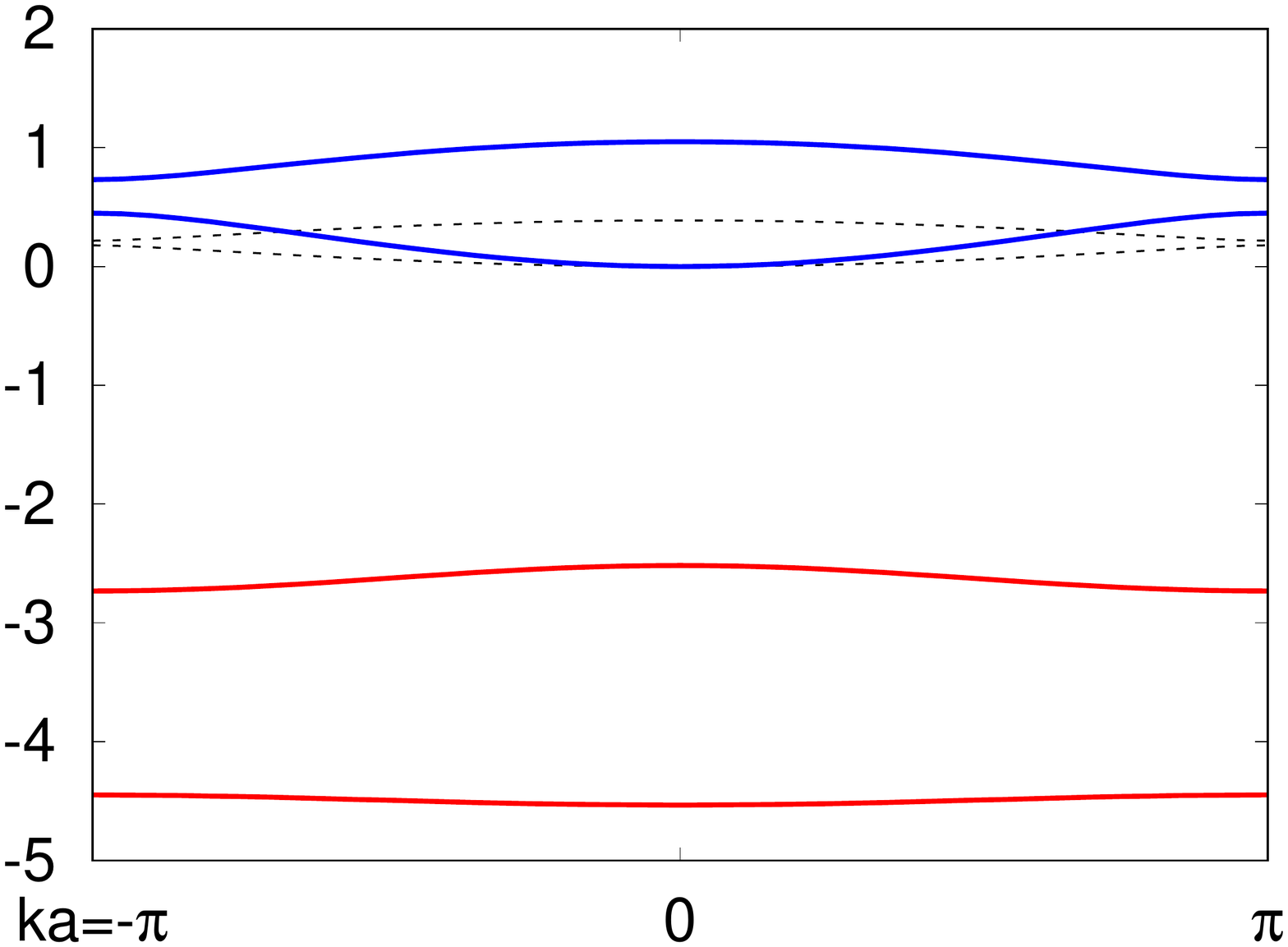}
  \caption{NiAs structure (left) reduced to a toy model comprising a
    one-dimensional chain of magnetic (B) and non-magnetic (A) atoms; there
    are two of the latter, A$_a$ and A$_b$ per unit cell. Energies are shown
    in the units of $t$ and $\Delta/t=1$, $-\epsilon_d/t=3$ was chosen.}
  \label{fig-03}
\end{figure}

Consider a 1D chain of alternating nonmagnetic (A) and magnetic (B)
atoms depicted in Fig.~\ref{fig-03} where only the nearest neighbours
couple (the amplitude being $t$). The single-orbital-per-site
tight-binding Hamiltonian assuming that the B-atom orbitals have
on-site energies $\epsilon_d\pm \Delta$ (where $2\Delta$ is the
exchange splitting) reads
\begin{equation}
  H_1(\Delta)=\left(\begin{array}{cccc}
    0 & t & 0 & te^{-ika} \\
    t & \epsilon_d+\Delta & t & 0 \\
    0 & t & 0 & t \\
    te^{ika} & 0 & t & \epsilon_d-\Delta 
    \end{array}
  \right)\label{eq-03}
\end{equation}
in the basis of Bloch states with momentum $k$ so that $ka$ ranging
from $-\pi$ to $\pi$ parametrises the BZ. 
  
The toy model described by $H_1$ can be treated analytically
(see Ref.~[\onlinecite{SI}], Sec.~1)
and focusing on the bands of dominantly A-atom orbital composition,
we observe a downfolded cosine band of width downscaled by factor
$\approx t/\epsilon_d$ in the large $|\epsilon_d|$ limit, i.e. remote
B-atom dominated bands, as the dashed black dispersion in Fig.~\ref{fig-03}
confirms. Atoms A$_a$ and A$_b$ interact only through the intermediate
(magnetic) atoms $B$ which suppresses their effective coupling.
There are two main observations to make at this
point. First, even if $\Delta\ll \epsilon_d$ there opens a gap in the
'VB states' at the BZ edge (to make the gap better visible, we chose
a larger value of $\Delta/\epsilon_d$ for the blue and red bands in
Fig.~\ref{fig-03}). %Size of the gap is parametrised by $\Delta$.
This allows for the insight that, inasmuch the atom A$_b$ is
sandwiched between spin-up (left) and spin-down neighbours (right), 
where the exchange coupling is $\Delta$ and $-\Delta$, their effect on
the A-band (blue in Fig.~\ref{fig-03} at the bottom right panel) does
not average out to zero. Next, an even more important insight concerns the
eigenstates of $H_1$ at $ka=\pm \pi$. 

At this point, we should point out that $H_1$ of~(\ref{eq-03}) in fact
only describes one of the two spin species; let us denote it as
up-spin and correspondingly, $H_{1,\uparrow}=H_1(\Delta)$. The two
states at $ka=\pm \pi$ split by nonzero $\Delta$ turn out to be
$(|a\rangle \pm |b\rangle)\otimes|\uparrow\rangle$ where $|a\rangle$
and $|b\rangle$ refer to orbitals of A$_a$ and A$_b$ atoms,
respectively. For the spin-down sector, $H_{1,\downarrow}=H_1(-\Delta)$ 
which leads to identical band structure as in Fig.~\ref{fig-03}
% (this is a common situation in antiferromagnets, see Appendix)
% maybe a comment on altermagnets here?
whose eigenstates are nevertheless not the same as for $H_{1,\uparrow}$. The
state degenerate with $(|a\rangle \pm |b\rangle)\otimes|\uparrow\rangle$
is $(|a\rangle \mp |b\rangle)\otimes|\downarrow\rangle$ and thus, we
arrive at the conclusion that, at the BZ edge, the VB states in our
toy model come in two pairs (split by the gap) and without loss of
generality, we now focus on the subspace spanned by the pair 
\begin{equation}
  (|a\rangle + |b\rangle)\otimes|\!\!\uparrow\rangle,
  (|a\rangle - |b\rangle)\otimes|\!\!\downarrow\rangle.
  \label{eq-02}  
\end{equation}
Unlike the pair $|\!\!\uparrow\rangle,|\!\!\downarrow\rangle$ (without
any orbital part), any linear combination of the two states in~(\ref{eq-02})
has a zero expectation value of transversal spin operators $\hat{\sigma}_x$,
$\hat{\sigma}_y$. This can also be restated as
$\langle\hat{\vec{\sigma}}\rangle || z$, or, easily generalised to the
statement that the states~(\ref{eq-02}) have the (expectation value
of) spin parallel to the magnetic moments of atoms B. In this way, the
{\em direction} of magnetic moments of the atoms remote in energy from
the VB top influences the current-carrying states close to the Fermi
energy. In the following, we denote the direction of spin in the basis
state $(|a\rangle + |b\rangle)\otimes|\!\!\uparrow\rangle$ by $\vec{L}$
and it can be understood as the N\'eel vector. \changed{In the following,
we explore this influence in the context of spin-orbit interaction; an
alternative pathway relies on spin disorder\cite{Baral:2022_a}
(as it occurs for example at
finite temperatures) and we outline an approach to it based on the coherent
potential approximation (CPA) in the Supplementary material. It provides
an alternative interpretation of 'magnetic blue shift'~\cite{Bossini:2020_a}
of the gap which does not rely on many-body effects.}

\subsection{Extension to MnTe crystal}

The previous argument can be extended to Te $p_x$, $p_y$ states which
form the VB top near A. To account for fine details of the band
structure (as explained in Ref.~\onlinecite{SI}, Sec.~2), we also
include the remote $p_z$ levels (in A, they are $\approx 3$~eV below
the VB top, see Fig.~2a) whose dispersion is dropped at this level of
approximation. \changed{Also note that the group of VB maxima close to
$\Gamma$ relies on Te $p_z$ orbitals as explained in the Supplementary
material.} We will measure energy from the VB top (as it appears
in the case of absent SOI) with $E_F$ denoting the Fermi energy and
use two copies of Eq.~(\ref{eq-01}) to describe the $k_{x,y}$-dependent
mixing of $p_x$, $p_y$ orbitals.

%We recall Eq.~(\ref{eq-01}) and take the
%parameters $a,b,c$ from the fit to ab initio calculations without SOI.

\begin{widetext}
Denoting the position of $p_z$ orbitals of tellurium by $e_z$
($e_z<0$, $|e_z/E_F|\gg 1$), the full description of the VB close to
A is provided by a block-diagonal $6\times 6$ matrix
\begin{equation}
  H_{kp}=\left(\begin{array}{cccccc}
    ak_x^2+bk_y^2+ck_z^2 & (a-b)k_xk_y&0 \\
    (a-b)k_xk_y & bk_x^2+ak_y^2+ck_z^2&0 \\
    0&0&e_z\\
    &&&ak_x^2+bk_y^2+ck_z^2 & (a-b)k_xk_y&0 \\
    &&&(a-b)k_xk_y & bk_x^2+ak_y^2+ck_z^2&0 \\
    &&&0&0&e_z
  \end{array}\right)
  \label{eq-06}
\end{equation}
%
% this is Eq.~3 in new/Documents/trav/In/19KdM/results/Manuscript_MnTe.pdf
%
and the first and second $3\times 3$ block is written in the
basis~(\ref{eq-02}) whereas inside the blocks, the basis vectors are
simply $|p_x\rangle,|p_y\rangle, |p_z\rangle$. Since the matrix~(\ref{eq-06})
does not explicitly depend on $\vec{L}$ (only its basis vectors are), we
arrive at the conclusion that (when SOI is ignored) the band structure does
not depend on the direction of Mn magnetic moments.

In the limit $|e_z|\to\infty$, the full model~(\ref{eq-06}) combined with
SOI breaks down into two decoupled $2\times 2$ blocks and since we 
now have a microscopic understanding of the basis, one which contains the
information about direction of Mn magnetic moments, the SOI can now be
evaluated. With finite $e_z$, the dependence of band splitting in A can
be better described as we explain in the following.

\end{widetext}

\begin{figure}
  \includegraphics[scale=0.4]{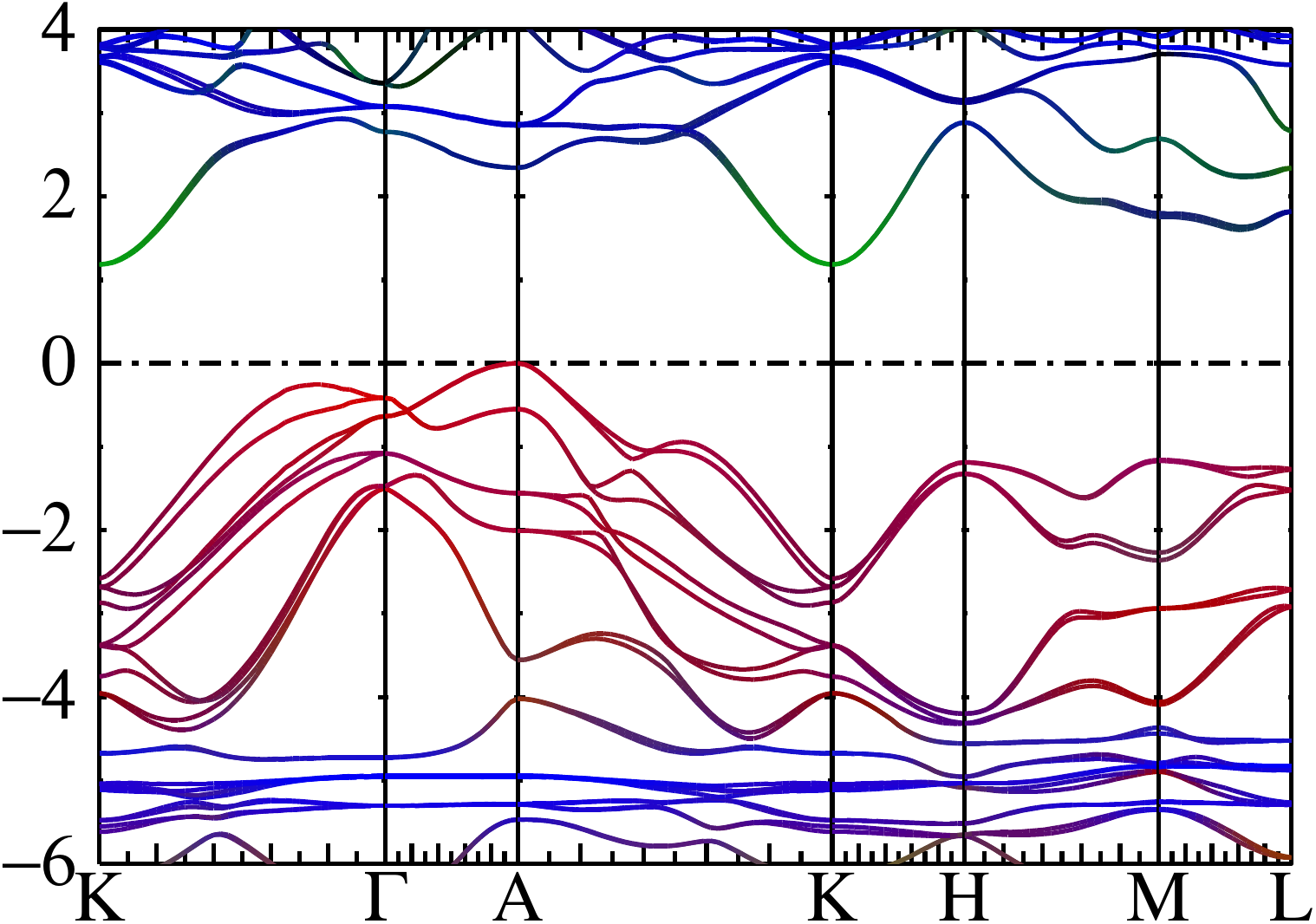}
  \caption{MnTe band structure for $\vec{L}||z$ (energies in eV).
    Colour coding: red Te, blue/green Mn.}
  \label{fig-04}
\end{figure}

\subsection{Spin-orbit interaction}

We are now in a position to explain the following behaviour of band
structure calculated by relativistic {\em ab initio} methods. In panel
(b) of Fig.~\ref{fig-02}, we could have already observed the bands
split by SOI and, compared to band widths, such splittings were
small \changed{(note that these splittings cause the shift of the VB
  top away from A, the point of high symmetry).}
Those calculations were done assuming $\vec{L}\parallel x$ and,
at this level of detail, depend only little on the direction of $\vec{L}$
as long as $\vec{L}\perp z$ which is compatible with MnTe being an
easy-plane material.\cite{Kriegner:2017_a} However, when
$\vec{L}\parallel z$ is assumed in 
calculations, see Fig.~\ref{fig-04}, band splittings become
sizable. Restricting our discussion to Te $p_x$, $p_y$ orbitals
combined into the states~(\ref{eq-02}), this
behaviour is linked to the directionality of $H_{so}=\lambda
\vec{l}\cdot\vec{\sigma}$ evaluated in the corresponding basis:
\begin{equation}
  H_{so,2\times 2}=\left(\begin{array}{cc}0&i\lambda\cos\theta\\
    -i\lambda\cos\theta&0\end{array}\right) 
  \label{eq-04}
\end{equation}
where $\vec{L}\cdot\hat{z}=\cos\theta$ and $\vec{l}$ is the orbital
angular momentum operator. Clearly, SOI projected to such a restricted
space becomes ultimately ineffective for the in-plane orientation of
magnetic moments where $\theta=\pi/2$ (taking into account also the
$|p_z\rangle$ orbital,\cite{SI} small splittings at A% Sec.2
in the in-plane configuration can nevertheless be also accounted for)
whereas for finite $e_z$, the full $6\times 6$ matrix of $H_{so}$ must
be considered instead of~(\ref{eq-04}). On the other hand, for out-of-plane
magnetic moments, the splitting at A seen in Fig.~\ref{fig-04} can be
directly compared to eigenvalues $\pm\lambda$ of $H_{so,2\times 2}$.

\section{Discussion and conclusions}

An effective model of the MnTe valence band around A point of the BZ
depends on the level of detail needed: Eq.~(\ref{eq-01}) is a meaningful
approximation to begin with but it cannot describe the band-dispersion
dependence on the direction of Mn magnetic moments; the six-band (or
four-band, corresponding to $|e_z|\to\infty$ limit) description using
Eq.~(\ref{eq-06}) combined with $H_{so}$ evaluated with respect to
basis~(\ref{eq-02}) times $|p_x\rangle,|p_y\rangle,|p_z\rangle$ is the
reasonable next step. On this level, the large sensitivity of the valence
band at A to magnetic moment orientation can be explained in terms of
zero matrix elements of $H_{so}$ between $(|a\rangle +
|b\rangle)\otimes|\!\!\rightarrow\rangle$ and
$(|a\rangle - |b\rangle)\otimes|\!\!\leftarrow\rangle$ where $a,b$ refer
to the two Te atoms within unit cell of MnTe. Zooming into the details
of the valence band smaller than $\sim 100$~meV would require adding
further terms such as those discussed on p. 8 of the supplementary information
to Ref.~[\onlinecite{Smejkal:2021_a}]; on this level of approximation,
phenomena such as the anomalous Hall effect or AMR can then likely be
successfully modelled.

Calculations in Fig.~\ref{fig-04} show that the splitting at A is
associated with reduction of band gap in agreement with DFT
calculations.\cite{Moseley:2022_a}  This implies that not only
angular-resolved photoemission (ARPES) could be used to confirm the
sensitivity of MnTe band structure to the orientation of Mn magnetic
moments but also optical absorption measurements should reveal
signatures of this effect. Such experiments could also confirm our
results concerning the competition of valence band maxima close to the
A and $\Gamma$ points of the Brillouin zone.

\section{Acknowledgements}

We acknowledge assistance of Swagata Acharya with QSGW calculations
and a preliminary KKR survey by Alberto Marmodoro; funding was
provided by grants 22-21974S and EU FET Open RIA No. 766566, M.v.S.
was supported by the DOE-BES Division of Chemical Sciences under
Contract No. DE- AC36-08GO28308 and P.E.F.Jr. acknowledges financial
support from the Alexander
von Humboldt Foundation, Capes (Grant No. 99999.000420/2016-06) and Deutsche
Forschungsgemeinschaft SFB 1277 (Project No. ID314695032, subprojects B05, B07
and B11).

% \begin{appendix}

% more appendices: 1) analytical results about H_1 and 'identical bands'
% for spin up and down, 2) d-term in Eq. 1 omitted (cmp to Koen's notes),
% 3) comparison with WIEN2k and QE calculations to GW

%\section{Analytical results concerning~(\ref{eq-03})}
% see Sec. 1 of the SI

%\section{More on spin-orbit term $H_{so}$}
% see Sec. 2 of the SI

%\end{appendix}


\begin{thebibliography}{99}
\bibitem{note1} Apart from the NiAs-type structure of MnTe,
  zinc-blende phase also exists which is AFM at low temperatures.
\end{thebibliography}
\end{document}